\documentclass[pra,twocolumn,showpacs,superscriptaddress,preprintnumbers,amsmath,amssymb,aps]{revtex4-1}
\usepackage{mathrsfs}
\usepackage{amsmath}
\usepackage{amssymb}
\usepackage{graphicx}
\usepackage{dcolumn}
\usepackage{bm}
\usepackage{subfigure}
\usepackage{color}
\usepackage{ifpdf}
\usepackage{epstopdf}
\usepackage{CJK}
\usepackage{verbatim}
\usepackage{bbm}
\usepackage{multirow}
\usepackage{booktabs}
\usepackage{color}
\usepackage[
pdfstartview=FitH,%
bookmarks=true,%
bookmarksopen=true,%
breaklinks=true,%
colorlinks=true,%
linkcolor=blue,anchorcolor=blue,%
citecolor=blue,filecolor=blue,%
menucolor=blue,pagecolor=blue,%
urlcolor=blue]{hyperref}
\usepackage{float}
\usepackage[toc,page,title,titletoc,header]{appendix}
\usepackage{ragged2e}
\usepackage{lipsum}

\usepackage{array}
\usepackage{booktabs}
\usepackage{graphics}
\usepackage{color}

\def\be{\begin{equation}}
	\def\ee{\end{equation}}
\def\bea{\begin{eqnarray}}
	\def\eea{\end{eqnarray}}
\def\nn{\nonumber}

\begin{document}
	\begin{CJK*}{GBK}{song}
\title{Transport of ultracold atoms in superpositions of S- and D-band states in a moving optical lattice}
\author{Zhongcheng Yu}
\affiliation{State Key Laboratory of Advanced Optical Communication System and Network, School of Electronics, Peking University, Beijing 100871, China}
\author{Jinyuan Tian}
\affiliation{State Key Laboratory of Advanced Optical Communication System and Network, School of Electronics, Peking University, Beijing 100871, China}
\author{Peng Peng}
\affiliation{State Key Laboratory of Advanced Optical Communication System and Network, School of Electronics, Peking University, Beijing 100871, China}
\author{Dekai Mao}
\affiliation{State Key Laboratory of Advanced Optical Communication System and Network, School of Electronics, Peking University, Beijing 100871, China}
\author{Xuzong Chen}
\affiliation{State Key Laboratory of Advanced Optical Communication System and Network, School of Electronics, Peking University, Beijing 100871, China}
\author{Xiaoji Zhou}\email{xjzhou@pku.edu.cn}
\affiliation{State Key Laboratory of Advanced Optical Communication System and Network, School of Electronics, Peking University, Beijing 100871, China}
\affiliation{Institute of Advanced Functional Materials and Devices, Shanxi University, Taiyuan 030031, China}
\date{\today}

\begin{abstract}
Ultracold atoms in a moving optical lattice with high controllability are a feasible platform to research the transport phenomenon. 
Here, we study the transport process of ultracold atoms at D band in a one-dimensional optical lattice, and manipulate the transport of superposition states with different superposition weights of S-band and D-band atoms.
In the experiment, we first load ultracold atoms into an optical lattice using shortcut method, and then accelerate the optical lattice by scanning the phase of lattice beams. 
The atomic transport at D band and S band is demonstrated respectively.
The group velocity of atoms at D band is opposite to that at S band.
By preparing superposition states with different superposition weights of D-band and S-band atoms, we realize the manipulation of atomic group velocity from positive to negative, and observe the quantum interference between atoms at different bands.
The influence of the lattice depth and acceleration on the transport process is also studied. Moreover, the multi-orbital simulations are coincident with the experimental results.
Our work sheds light on the transport process of ultracold atoms at higher bands in optical lattices, and provides a useful method to manipulate the transport of atomic superposition states.
\end{abstract}
		
		
\maketitle
\end{CJK*}
	
\section{Introduction}


The transport phenomenon has been attracting tremendous efforts in recent years \cite{RevModPhys.83.407,RevModPhys.91.035001}, which occurs in a system that particles are driven by an external force and move in a periodic potential \cite{PhysRevLett.103.090403,PhysRevA.82.013633}. 
The transport process is studied in the fields of electronic materials \cite{science.1174736,science.1148047}, trapping ions \cite{Kielpinski2002,PhysRevLett.107.023002}, and ultracold atoms \cite{PhysRevA.71.053606,BAKHTIARI2006223,PhysRevLett.98.120403}.

The ultracold atoms in optical lattices, due to high controllability, are widely applied to simulating the physics of condensed matter \cite{RevModPhys.80.885,RevModPhys.91.015005,Li_2016}. 
For instance, novel physical phases \cite{science.aaf6689,PhysRevA.74.013607,PhysRevLett.111.185301,Jotzu2014} and dynamical mechanisms of atomic superfluid \cite{RevModPhys.78.179,PhysRevLett.95.170404,PhysRevA.87.063638,Tarnowski2019} in optical lattices are observed with various manipulation technologies \cite{Zhou_2018,science.aad4568}, and atomic circuits and atomic qubits are achieved in optical lattices \cite{PhysRevLett.103.140405,PhysRevA.104.L060601}.
Among them, higher orbital physics in optical lattices has attracted much attention, such as the recoagulation of the P-band Bosons in a hexagonal lattice \cite{PhysRevLett.126.035301,Wang2021}, the achievement of Ramsey interferometry between S band and D band \cite{Hu2018}, the observation of atomic scattering at D band \cite{PhysRevA.104.033326}, and the unconventional superfluid order at F band \cite{PhysRevLett.106.015302}. 

Furthermore, optical lattices are also an effective platform to study transport phenomena \cite{PhysRevA.84.033638,PhysRevA.74.013403,PhysRevA.72.043618,Holthaus_2000}, such as the research on long-range transport \cite{Middelmann_2012,PhysRevLett.122.010402,PhysRevLett.104.200403}, and the effects of Bosonic and Fermionic statistics on atomic transport \cite{PhysRevA.85.041601}.
The moving optical lattice is a powerful tool to study the transport process.
The inertia force produced by the moving optical lattice has the advantages of a large adjustable range \cite{PhysRevLett.101.230801,PhysRevLett.98.120403,PhysRevLett.103.090403} and a controllable direction \cite{Tarnowski2019}.
And the transport process in moving optical lattices is widely applied in quantum simulation and quantum precision measurement. For example, it is applied in detecting topological properties\cite{science.abm6442}, and improving the integration time of atomic gravimeters \cite{PhysRevLett.101.230801,PhysRevA.88.031605}.
However, there are few researches on the transport of atoms at higher bands and superposition states of different bands in optical lattices.

In this work, the transport process of ultracold atoms at D band is observed in a moving one-dimensional optical lattice for the first time, and we perform a transport manipulation of superposition states with S-band and D-band atoms, which realizes the change of atomic group velocities from positive to negative.
By using our proposed shortcut method \cite{Zhou_2018}, we load atoms from a harmonic trap into the optical lattice with different states within tens of microseconds.
Then we perform the transport process of D band atoms in the moving one-dimensional optical lattice, and observe the group velocity of D-band atoms is opposite to that of S-band atoms.
By their transport characteristic, we modulate the transport process of superposition states with different superposition weights of D-band and S-band atoms, and the group velocity from positive to negative is observed. 
We compare the transport of superposition states to classical mixtures, and observe the quantum interference between atoms at different bands.
Further, we study the influence of the lattice depth and acceleration on transport process.
Using the multi-orbital method \cite{PhysRevA.104.L060601}, the atomic group velocities with different parameters are calculated, and the calculations agree with the experimental results.
This work paves the way for the manipulation of transport process with ultracold atoms at higher bands in optical lattices, and is helpful to research the transport of atomic superposition states. 

\begin{figure*}[ht]
	\includegraphics[width=0.7\textwidth]{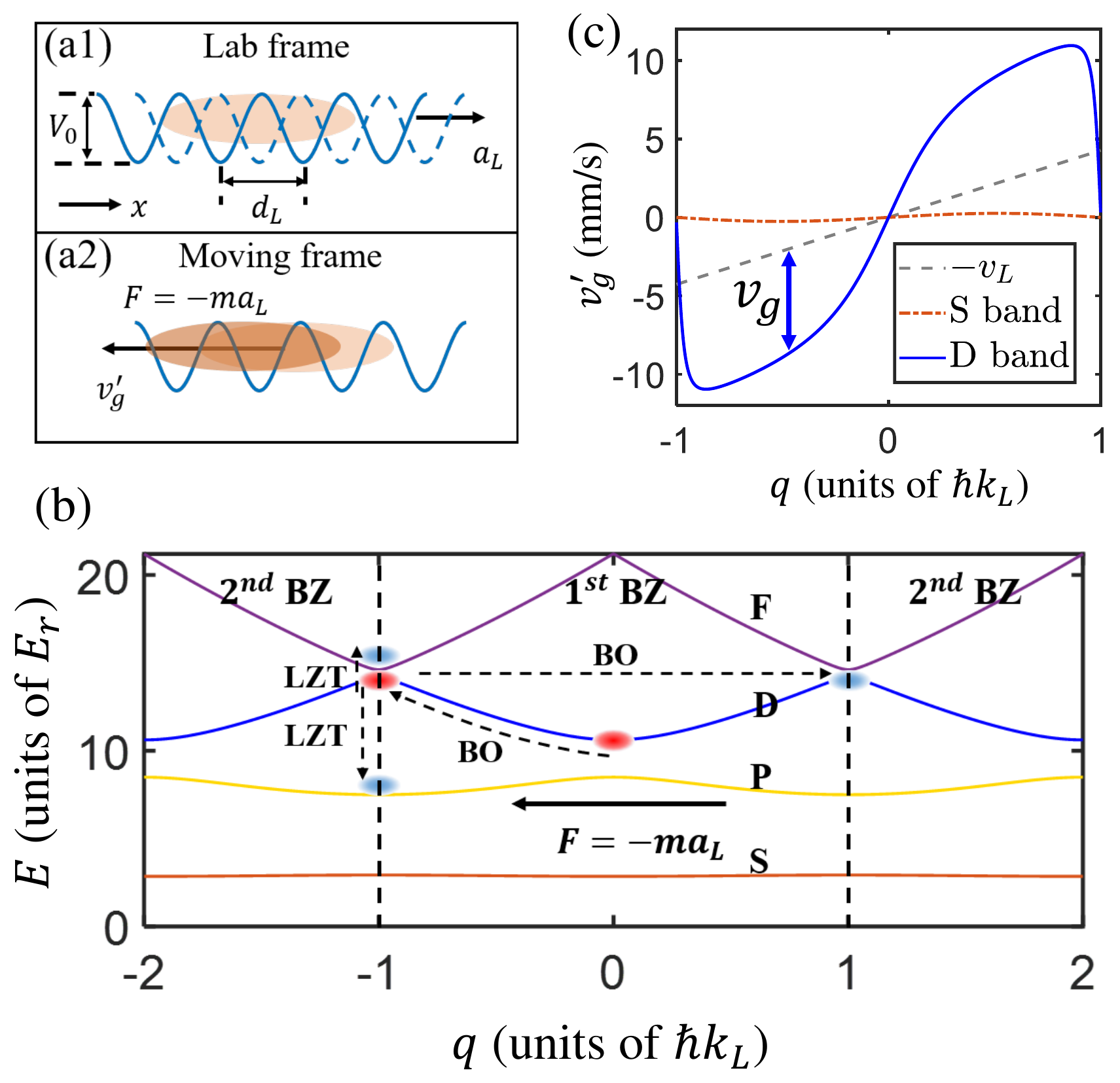}
	\caption
	{\textbf{Transport process of atoms in a moving optical lattice.}
		(a1) and (a2) show the ultracold atoms in a moving optical lattice in the laboratory frame and moving frame, respectively. In (a1), the solid blue line represents the optical lattice, which is moving with acceleration $a_L$. The dashed blue line shows the motion of optical lattice. $d_L$ is the lattice constant, and $V_0$ is the lattice depth. The arrow $x$ denotes the position coordinate.
		In (a2), $F=-ma_L$ is the inertia force, and $v_g'$ is the group velocity of atoms in moving frame.
		(b). The transport process of D-band atoms in moving frame. The orange, yellow, blue, and purple lines (from bottom to top) represent the dispersion relation of S, P, D, and F bands in quasi-momentum space, respectively. The lattice depth is $10$ $E_r$. The dashed arrows denote Bloch oscillation (BO) and Landau-Zener tunnel (LZT). The red ellipses denote the atoms performing Bloch oscillation at a single band. 
		The blue ellipses denote the atoms which perform Landau-Zener tunneling to other bands or scattering to another edge of the first Brillouin zone. 
		The dashed vertical lines show the edges of the first Brillouin zone.
		(c). The group velocity of atoms in the laboratory frame and moving frame. The dash-dotted orange line and solid blue line represent $v_g'$ of atoms at S band and D band in the moving frame, and the dashed line denotes the velocity $-v_{L}=q/m$ of optical lattice. The distance between $-v_{L}$ and $v_g'$ of S band and D band shows the group velocity $v_g$ in the laboratory frame.
	}\label{fig:Fig_principle}
\end{figure*}

This paper is organized as follows. In Sec. II, we describe the transport process of S-band and D-band atoms in the moving lattice, and the method of numerical calculation is given. 
Our experimental setup and time sequences are represented in Sec. III A.
In Sec. III B, we introduce shortcut method used to prepare atoms into target states. 
The experimental results of transport process in a single band and superposition states of atoms at S band and D band are shown in Sec. IV A and Sec. IV B, respectively. 
And the influence of the lattice depth and acceleration on the transport process is studied in Sec. IV C. Finally, we give the discussion and conclusions in Sec. V.

\section{Transport process in a moving optical lattice and theoretical simulation}

\subsection{Transport process in a moving optical lattice}
We study the transport process of ultracold atoms in a moving one-dimensional optical lattice.
As shown in Fig.\ref{fig:Fig_principle} (a), in the laboratory frame (Lab frame), the optical lattice is moved with acceleration $a_L$, and the Hamiltonian of this system is:
\begin{eqnarray}
	\label{H_ao}
	\hat{H}=\frac{\hat {p}^2}{2m}+\frac{1}{2}V_0 \cos(2k_L(x-\frac{1}{2}a_Lt^2)),
\end{eqnarray}
where $\hat {p}$ is the momentum operator, $m$ is the mass of an atom, $V_0$ is the lattice depth, $k_L=2\pi/\lambda$ is the wave vector of optical lattice, and $\lambda$ is the wavelength of the lattice beams.
The Hamiltonian Eq.(\ref{H_ao}) can be considered as atoms applied by an inertia force in the moving lattice frame. To analyze the motion of atoms, we first consider the atomic motion in moving frame, and then transform the reference system to Lab frame by adding the velocity of optical lattice.

Fig.\ref{fig:Fig_principle} (b) shows the motion of atoms at D band in moving frame.
Initially, the atoms distribute at the center of the first Brillouin zone (BZ). Due to the inertia force $F=-ma_L$, atoms perform Bloch oscillation (BO) and the quasi momentum $q$ linearly changes with $t$ \cite{PhysRevLett.87.140402,PhysRevLett.82.2022}:
\begin{eqnarray}
	\label{BO}
	q(t)=Ft=-ma_Lt.
\end{eqnarray}
During Bloch oscillation, some atoms perform Landau-Zener tunneling (LZT) to other bands.
To explain the transport process, firstly we only consider Bloch oscillation, and the LZT and BO will be considered comprehensively in subsection B. For Bloch oscillation, the group velocity in moving frame $v_g'(q(t))$ is determined by the dispersion relation $E_{\alpha}(q)$:
\begin{eqnarray}
	\label{vgt}
	v_g'(q)=\hbar^{-1}{\rm d}E_{\alpha}(q)/{\rm d}q,
\end{eqnarray}
where $\alpha$ represents the band and can be chosen as $S,P,D,F...$ band.

Connecting Eqs. (\ref{BO}) and (\ref{vgt}), the group velocity in moving frame is given:
\begin{eqnarray}
	\label{vgtt}
	v_g'(t)=\hbar^{-1}\frac{{\rm d}E_{\alpha}(q)}{{\rm d}q}|_{q=-ma_Lt}.
\end{eqnarray}
The solid blue line and dash-dotted orange line in Fig. \ref{fig:Fig_principle} (c) represent $v_g'(t)$ for atoms at D band and S band, respectively.

Next, we transform the reference system to Lab frame by adding the velocity of optical lattice $v_L(t)$ to the velocity of atoms. And the group velocity of atoms $v_g(t)$ is given:
\begin{eqnarray}
	\label{vg}
	v_g(t)=v_g'(t)+v_L(t).
\end{eqnarray}
Using Eq.(\ref{BO}), the velocity of optical lattice $v_L$ can be written as:
\begin{eqnarray}
	\label{vl}
	v_L=a_Lt=-q/m.
\end{eqnarray}
In Fig. \ref{fig:Fig_principle} (c), the dashed line denotes the velocity $-v_L=q/m$. 
$v_g'(t)$ of S band and D band are on the opposite sides of $-v_L$.
Hence, in Lab frame, the atoms at S band and D band will obtain the group velocity in the opposite direction. This result is different from the situation that atoms are driven by an external force in a stationary lattice, where the atoms at S band and D band will get the group velocity in the same direction.

\subsection{Multi-orbital calculation method}
Using the multi-orbital simulation method \cite{PhysRevA.104.L060601}, we consider Landau-Zener tunneling and Bloch oscillation comprehensively, and calculate the motion of atoms.

To begin with, we divide the time $t$ into ${\rm d}t$, and project the Hamiltonian Eq.(\ref{H_ao}) into the momentum states $\left | n \right \rangle=e^{2ink_L x}$, where $n=0,\pm1,\pm2...$, and $x$ is the space coordinate position.
Ignoring the interaction, the instantaneous Hamiltonian is written as \cite{PhysRevA.104.L060601}:
\begin{eqnarray}
	\label{H_ins}
	H_{nn'}=\frac{\hbar^2 k_L^2}{2m}(2n)^2\delta_{n,n'}\\ \nn
	+\frac{1}{4}V(t)\times(\delta_{n,n'+1}+\delta_{n,n'-1}),
\end{eqnarray}
where $n,n'=0,\pm1,\pm2,...$, and $V(t)$ is the potential energy of the instantaneous Hamiltonian. 
Solving the eigenvalue equation of the instantaneous Hamiltonian, the instantaneous eigenvalue $\varepsilon_\alpha(t)$ is gotten.
By implementing all the instantaneous evolution operators $e^{-i\varepsilon_\alpha(t) {\rm d}t/\hbar}$ to the initial state $\left | \alpha, n=0 \right \rangle$, the state at time $t$ is given:
\begin{eqnarray}
	\label{psi_f}
	\left | \psi(t) \right \rangle=\prod_{t}e^{-i\varepsilon_\alpha(t) {\rm d}t/\hbar}\left | \alpha, n=0 \right \rangle.
\end{eqnarray}
The group velocity $v_g(t)$ at time $t$ is:
\begin{eqnarray}
	\label{vg_cal}
	v_g(t)=\left \langle \psi(t) \right | \frac{\hat {p}}{m}\left | \psi(t) \right \rangle,
\end{eqnarray}
where momentum operator $\hat {p}$ can be written as $\sum_{n}2n\hbar k_L\left |n \right \rangle\left \langle n \right |$.

\section{Experimental description}

\subsection{Experimental setup and Sequence}
\begin{figure}
	\includegraphics[width=0.45\textwidth]{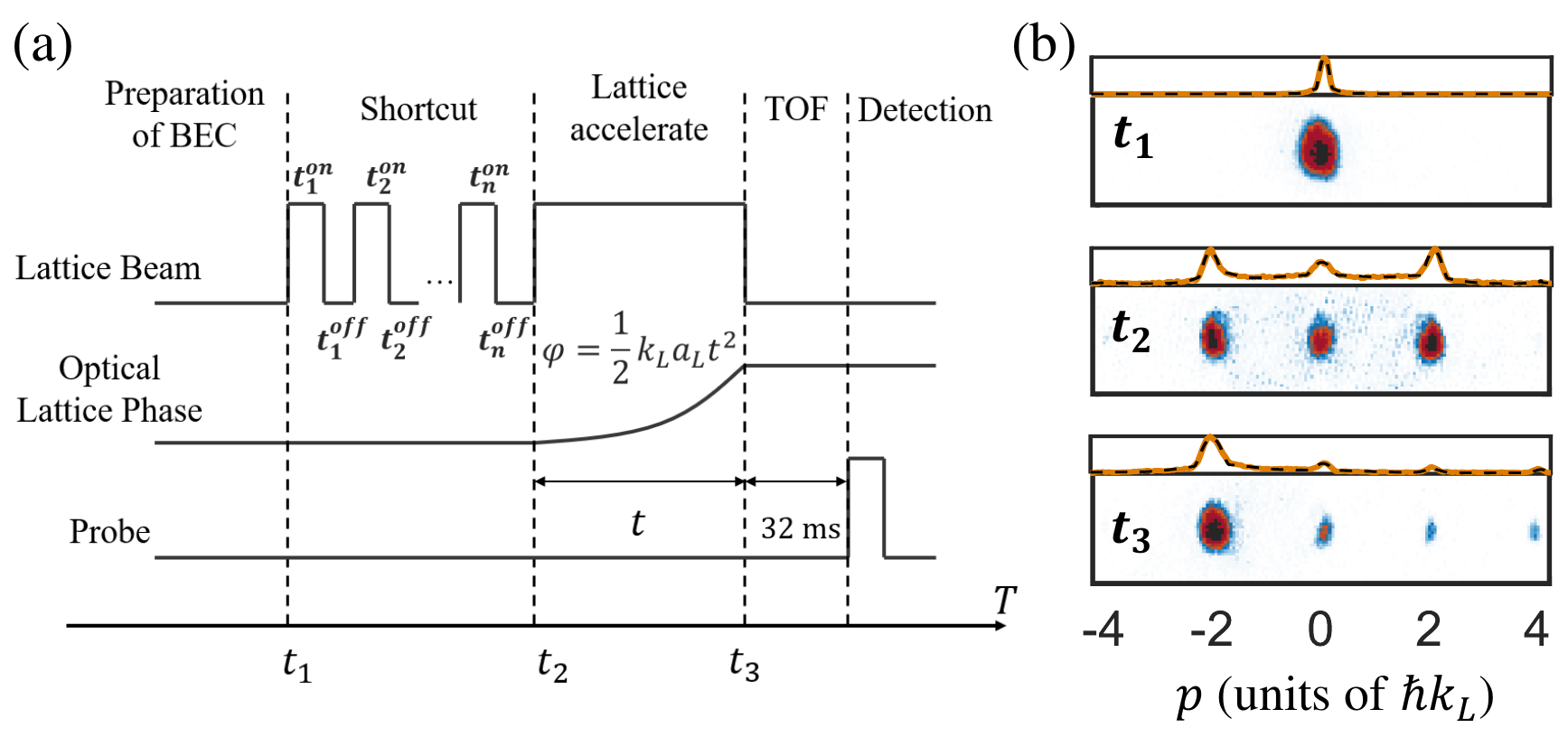}
	\caption{\textbf{The experimental sequence and typical images.} (a)The experimental sequence. The horizontal axis denotes the experimental time, and there are several stages of the experiment. The first stage is the preparation of BEC. In the second stage, shortcut method is used to prepare the atomic states in the optical lattice. $t^{\rm on(off)}_i$ shows the pulse sequence of shortcut method. In the third stage, the optical lattice is accelerated by changing the phase of lattice beam. After the time of flight 32 ms, the absorption imaging is taken to detect the atoms. $t_1,t_2,t_3$ are three typical moments: the BEC, the atoms loaded into the optical lattice, and the atoms after accelerating. (b) shows the typical images corresponding to the moment $t_1,t_2,t_3$. The above curves show the fitting of images. The horizontal axis represents the momentum $p$.
	}\label{fig:Fig_device}
\end{figure}

Our experiment starts with a Bose Einstein condensate (BEC) of $^{87}$Rb with $2\times 10^5$ atoms which are prepared in a hybrid trap.
The experimental setup has been described in our previous works \cite{PhysRevLett.126.035301,PhysRevLett.121.265301}. 
Then, we use shortcut method to load atoms into the one-dimensional optical lattice (see Subsection B), and move the lattice with the acceleration $a_L$ for time $t$, as shown in Fig. \ref{fig:Fig_device}(a). 
The one-dimensional optical lattice is composed of a 1064 nm laser beam and its reflected beam, and an electro-optic modulator (EOM, Thorlabs EO-PM-NR-C2) is placed in the optical path of the reflected beam. The EOM produces an additional phase $\varphi(t)$ to the reflected beam. Hence, the lattice potential $V_L(x,t)$ can be written as:
\begin{eqnarray}
	\label{lattice}
	V_L(x,t)=\frac{1}{2}V_0 \cos(2k_Lx-2\varphi(t)).
\end{eqnarray}
By controlling the voltage of EOM, the phase $\varphi(t)$ can be scanned.
To accelerate the optical lattice with $a_L$, as shown in Fig. \ref{fig:Fig_device}(a), the phase $\varphi(t)$ is scanned as:
\begin{eqnarray}
	\label{lattice_phase}
	\varphi(t)=\frac{1}{2}a_Lk_Lt^2.
\end{eqnarray}
After evolving for time $t$, we turn off the lattice beam and take absorption imaging with the time of flight (TOF) $t_{\rm TOF}=32 ~{\rm ms}$ to detect the transport process.

Fig.\ref{fig:Fig_device}(a) shows the experimental sequence, where time $t_1,t_2,t_3$ denote the stage of BEC, the stage of atoms loaded into the optical lattice, and the stage after accelerating, respectively. Fig.\ref{fig:Fig_device}(b) shows the typical images at these three moments. At $t_1$, the atoms condensate with zero momentum. After the pulse sequence of shortcut method at $t_2$, the atoms are loaded into the optical lattice, and symmetrically distribute as several momentum peaks. The population proportion of each momentum peak is determined by the proportion and phase of atoms at each band, but the group velocity of atoms at $t_2$ is always zero. After accelerating, at $t_3$, the distribution of atoms at each momentum peak changes, and the atoms have a non-zero group velocity. 
To deal with the absorption images, we calibrate the distance of momentum $2\hbar k_L$ in the images by fitting the center of two adjacent atom clouds. By the calibration, we can obtain the momentum of atoms in the images.

In the experiment, limited by EOM control voltage, the maximum reachable phase of the laser beam is fixed.
Hence, the range of measured time is limited for a chosen $a_L$.
We choose $a_L$ around several times gravitational acceleration, where the transport process is obvious.
One feasible method to increase the measured time is to use additional EOMs.

\subsection{Shortcut method to load atoms into the target states of an optical lattice}
We use shortcut method, due to its simplicity and efficiency \cite{Zhou_2018}, to prepare atoms into target states in an optical lattice. 
The basic idea of shortcut method is to continually turn on and off the optical lattice to modulate the atomic state, as shown in Fig. \ref{fig:Fig_device}(a). In our experiment, the initial state $\left | \psi_i \right \rangle$ is the ground state of BEC in a harmonic trap. After several pulses, the final state of atoms $\left | \psi_f \right \rangle$ is written as \cite{Zhou_2018,PhysRevA.104.033326}:
\begin{eqnarray}
	\label{state_afterpluses}
	\left | \psi_f \right \rangle =  \prod \limits_{j=n_p}^1 \hat{U}^{\rm off}(t^{\rm off}_{j}) \hat{U}^{\rm on}(t^{\rm on}_j) \times \left | \psi_{i} \right \rangle,
\end{eqnarray}
where $n_p$ is the number of pulses, $\hat{U}^{\rm on(off)}$ is the evolution operator of atomic state when the optical lattice is on (off), and $t^{\rm on(off)}_{j}$ is the evolution time of the $j^{\rm th}$ pulse. 

Through designing the sequence $t^{\rm on(off)}_{j}$, we can transfer the BEC from the Harmonic trap into our target state $\left | \psi_a \right \rangle$ with a high fidelity in tens of microsecond. The fidelity $\eta$ describes the efficiency of state preparation and is defined as \cite{Zhou_2018,PhysRevA.104.033326}:
\begin{eqnarray}
	\label{shortcut_fieldty}
	\eta=|\left \langle \psi_f|\psi_a \right \rangle |^2.
\end{eqnarray}

In the experiment, the target state $\left | \psi_a \right \rangle$ is the state of a single band or the superposition of S-band state $\left | S \right \rangle$ and D-band state $\left | D \right \rangle$:
\begin{eqnarray}
	\label{aimed_state}
	\left | \psi_a \right \rangle=\gamma_1 \left | S \right \rangle+
	\gamma_2 \left | D \right \rangle,
\end{eqnarray}
where $|\gamma_1|^2+|\gamma_2|^2=1$. For the state $\left | S \right \rangle$ or $\left | D \right \rangle$, the quasi momentum is zero, and its relative phase is zero.

\begin{table}[htp]
	\caption{The pulse sequences used to prepare atoms into the target state $\left | \psi_a \right \rangle$ of the one-dimensional optical lattice with $V_0=10 ~E_r$.}
	\label{tab:shortcut_seq}
	\begin{tabular}{cccccccccc}
		\hline
		\specialrule{0em}{1pt}{1pt}
		{$t^{\rm on}_1$($\rm \mu s$)}    & {$t^{\rm off}_1$}    & {$t^{\rm on}_2$}    & {$t^{\rm off}_2$}    & {$\gamma_1$} & {$\gamma_2$}  & {$\eta^{T}$} & {$\eta^{E}$}  \\
		\hline
		\specialrule{0em}{1pt}{1pt}
		{$10$}    & {$32$}    & {$26$}    & {$10$}  & {$1$} & {$0$}  & {$0.9999$} & {$0.994$} \\
		{$2$}    & {$30$}    & {$0$}    & {$0$}  & {$\sqrt{0.8}$} & {$\sqrt{0.2}$}  & {$0.9994$} & {$0.974$} \\
		{$14$}    & {$47$}    & {$11$}     & {$34$}  & {$\sqrt{0.6}$} & {$\sqrt{0.4}$}  & {$0.9996$} & {$0.964$} \\
		{$21$}    & {$2$}    & {$31$}     & {$39$}  & {$\sqrt{0.3}$} & {$\sqrt{0.7}$}  & {$0.9994$} & {$0.956$} \\
		{$22$}    & {$43$}    & {$61$}    & {$39$}   & {$0$} & {$1$}  & {$0.9998$} & {$0.991$} \\
		\hline
	\end{tabular}	
\end{table}

Table \ref{tab:shortcut_seq} shows the pulses used in the experiment, where $V_0=10 ~E_r$ and $E_r=\frac{\hbar^2 k_L^2}{2m}$. The theoretical fidelity $\eta^{T}$ for each sequence is near 1, and $\eta^{E}$ is the experimental fidelity.

\section{Experimental results}

\subsection{Transport process of atoms distributing at a single band}
\begin{figure}
	\includegraphics[width=0.48\textwidth]{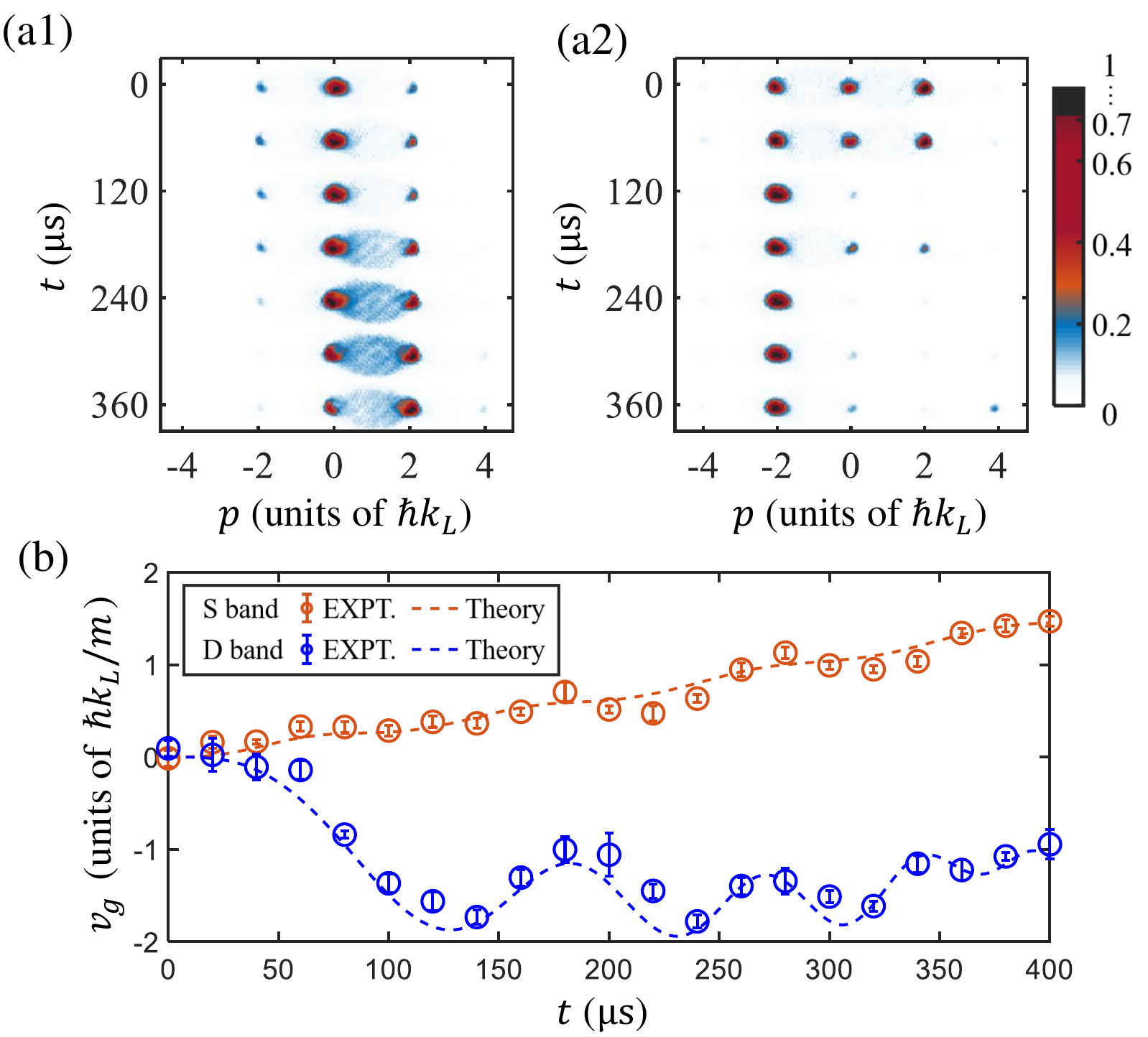}
	\caption{\textbf{The transport process of atoms at S band and D band.}
		(a1) and (a2) are the absorption images of transport process of S-band and D-band atoms, respectively. The horizontal axis represents momentum $p$, and the vertical axis is time $t$. The color denotes the normalized density of atoms.
		(b) The blue (lower) data and orange (upper) data show the evolution of group velocity $v_g$ of D-band atoms and S-band atoms. The circles are the experimental data, and the dashed lines are the simulations. The errorbar denotes the standard error of five measurements. The parameters in the figure: $V_0=10 ~E_r$, $a_L=15 ~{\rm m/s^2}$.
	}\label{fig:Fig_result1}
\end{figure}
By shortcut method and the moving lattice, we study the transport process of atoms at S band and D band. Fig. \ref{fig:Fig_result1} (a1) and (a2) show typical images of S-band and D-band atoms, where $V_0=10 ~E_r$, $a_L=15 ~{\rm m/s^2}$. The vertical axis of Fig. \ref{fig:Fig_result1} (a1) and (a2) represent time $t$, and the horizontal axis represents the atomic momentum $p$. The color denotes the normalized density of atoms. 
In the images, atoms distribute around momentum peaks $p=\pm 2\xi\hbar k_L$ ($\xi=0,\pm 1,\pm 2, ...$), and the number of atoms at each momentum peak changes over time.
In moving frame, the velocity of each momentum peak increases as $q/m$, and in Lab frame the velocity becomes $q/m+a_Lt=0$.
Hence, the center of each momentum peak remains unchanged, and the change of group velocity is reflected in the transfer of atom population at each momentum peak.

For the atoms initially at S band, they mainly distribute in the central zero-order momentum peak, while a small part of them symmetrically distribute in the positive and negative first-order momentum peaks, as shown in Fig. \ref{fig:Fig_result1} (a1). Within the time range of this experiment, the atoms at S band gradually move towards positive momentum peaks. The scattering halos between two momentum peaks may be from the collisions of different momentum components \cite{Chatelain_2020}. 
By calculating the average momentum, the group velocity of atoms at S band is acquired, as the orange (upper) data in Fig. \ref{fig:Fig_result1}(b) shows. 
$v_g$ of S-band atoms rises with small oscillations, which agrees with the theoretical dashed line. 

For the atoms initially at D band, they symmetrically distribute in the three central momentum peaks, as shown in Fig. \ref{fig:Fig_result1} (a2). There are much more atoms distributing in the positive and negative first-order momentum peaks than that at S band, which is due to the higher energy of atoms at D band.
As time $t$ increases, the atoms at D band gradually move towards negative momentum peaks.
Using the same method as that for S-band atoms, we calculate the group velocity of D-band atoms, which changes from zero to negative and then gradually oscillates over time, as the blue (lower) points in Fig. \ref{fig:Fig_result1}(b) show. The experimental results are consistent with the analysis in Sec. II A that the group velocity of D-band atoms obtained from the moving lattice is opposite to the group velocity of S-band atoms.

\begin{figure}
	\includegraphics[width=0.4\textwidth]{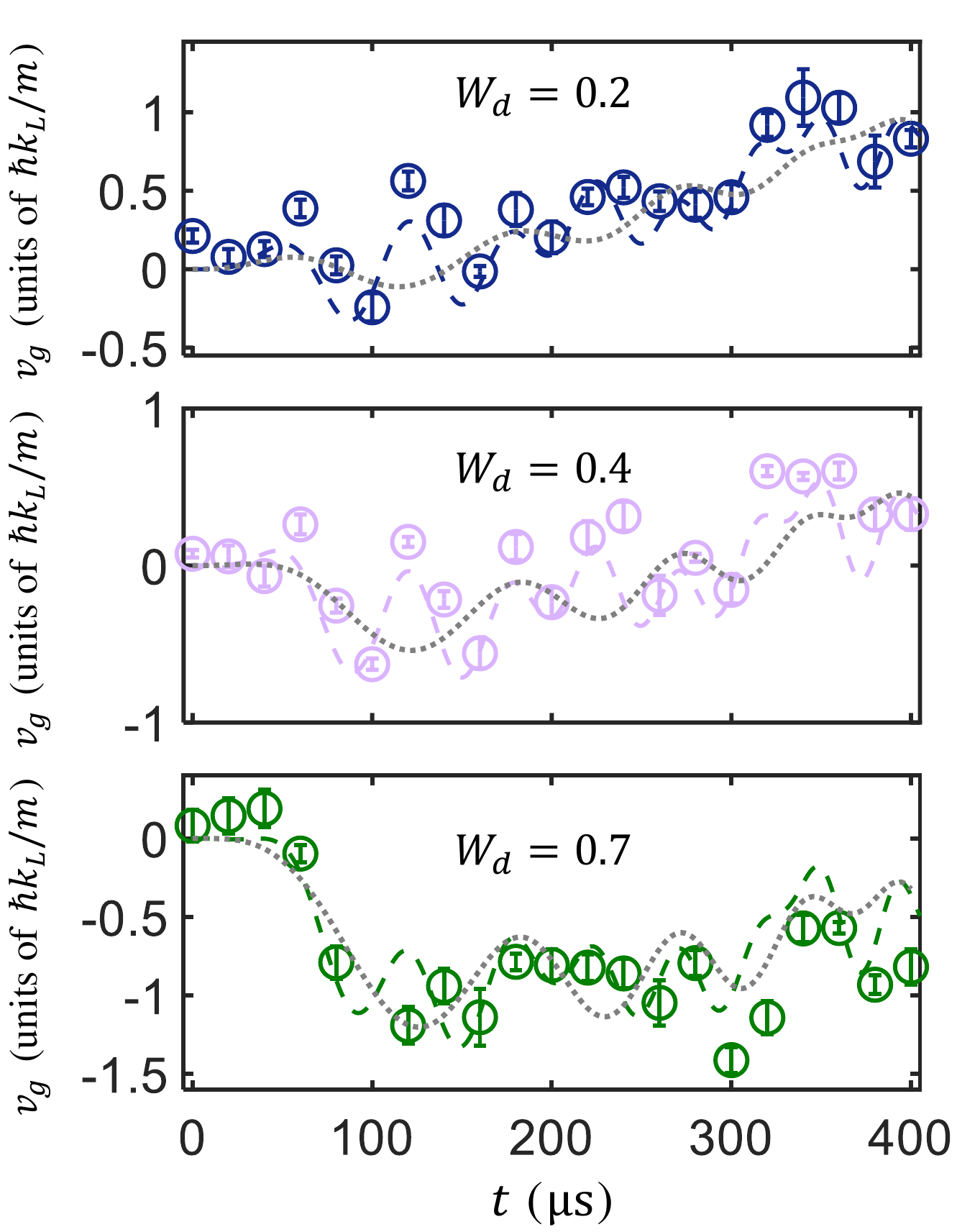}
	\caption{\textbf{The group velocity $v_g$ for the different superposition weights of D band $W_d$ over time.} The top, middle, and bottom figures correspond to the case of $W_d=$0.2, 0.4, and 0.7, respectively. The dashed lines denote the theoretical superposition group velocity curves, calculated by the multi-orbital simulation method. The gray dotted lines denote the theoretical classical group velocity curves, calculated by weighted averaging of the S-band and D-band group velocity. The errorbar denotes the standard error of five measurements. The parameters in the figure: $V_0=10 ~E_r$, $a_L=15 ~{\rm m/s^2}$. 
	}\label{fig:Fig_result2}
\end{figure}

\subsection{Transport process of atomic superposition states}
Since the group velocity of S-band atoms and D-band atoms are opposite, by preparing superposition states with different superposition weights of atoms at D band and S band, the group velocity obtained from the accelerating optical lattice can be modulated.
The superposition weight of D-band atoms $W_d$ is defined as:
\begin{eqnarray}
	\label{Pd}
	W_d=\left \langle \hat{N}_d \right \rangle/\left \langle \hat{N} \right \rangle,
\end{eqnarray}
where $\left \langle \hat{N}_d \right \rangle$ is the expectation value of D-band atom number operator, and $\left \langle \hat{N} \right \rangle$ is the expectation value of the total atom number operator.

Using shortcut method, we design the sequences to prepare the atomic superposition states in optical lattice with different $W_d$.
Fig. \ref{fig:Fig_result2} shows the group velocities obtained from the moving optical lattice with different $W_d$. 
The top figure of Fig. \ref{fig:Fig_result2} shows the transport process of $W_d=0.2$.
The group velocity gradually rises, but the increasing rate is slower than that only distribute at S band.
As $W_d$ increases, the group velocity of atoms gradually decreases.
The middle figure of Fig. \ref{fig:Fig_result2} shows the atomic group velocity with $W_d=0.4$, which oscillates around zero.
When $W_d$ reaches 0.7, as shown in the bottom figure of Fig. \ref{fig:Fig_result2}, the group velocity first changes from zero to negative and then gradually oscillates over time, like the behavior of D-band atoms.

Due to the coherence, the transport of superposition states is different from classical mixtures.
The difference comes from the quantum interference between atoms at different bands \cite{PhysRevA.82.065601}, and the group velocity of atomic superposition states in our experiment performs more oscillations than that of classical mixtures.
In Fig. \ref{fig:Fig_result2}, the dashed lines are the theoretical group velocity of the superposition states, which are calculated by the multi-orbital simulation method.
The gray dotted lines are the theoretical group velocity of classical mixtures, which are calculated by the weighted averaging of S-band and D-band group velocity.
To compare the deviation of two theory curves from the experiments, we define the dimensionless root mean squared error between the experimental results and theoretical simulations \cite{Bernstein1999}:
\begin{eqnarray}
	\label{sigma}
	\sigma=\frac{m}{\hbar k_L}\sqrt{\frac{1}{N_t}\sum_{i=1}^{N_t}(v_g^e(t_i)-v_g^t(t_i))^2},
\end{eqnarray}
where $v_g^e(t_i)$ and $v_g^t(t_i)$ are the experimental and theoretical group velocities at time $t_i$, and $N_t$ is the number of measured moments. 
\begin{table}[htp]
	\caption{The deviation of superposition curve and classical curve from the experimental results, when $V_0=10 ~E_r$ and $a_L=15 ~{\rm m/s^2}$.}
	\label{tab:error}
	\begin{tabular}{ccc}
		\hline
		\specialrule{0em}{1pt}{1pt}
		{~~~$W_d$~~~}    & {~~~$\sigma_s$~~~}    & {~~~$\sigma_c$~~~} \\
		\hline
		\specialrule{0em}{1pt}{1pt}
		{$0.2$}    & {$0.247 \pm 0.026$~}    & {~$0.301 \pm 0.017$} \\
		{$0.4$}    & {$0.266 \pm 0.013$~}    & {~$0.339 \pm 0.016$} \\
		{$0.7$}    & {$0.336 \pm 0.008$~}    & {~$0.342 \pm 0.011$} \\
		\hline
	\end{tabular}	
\end{table}
Table \ref{tab:error} shows the dimensionless root mean squared error of superposition curve $\sigma_s$ and classical curve $\sigma_c$, where the uncertainty is calculated by the standard error of five measurements. 
All the $\sigma_s$ is smaller than $\sigma_c$ which means the deviation of the superposition curve from the experiment results is smaller than that of the classical curve. Hence, the experimental results are more consistent with the curve of atomic superposition states.

When $W_d=0.2,0.4$, the experimental uncertainty is smaller than the difference between superposition curve and classical curve. When $W_d=0.7$, the difference between superposition curve and classical curve is not obvious, and it may be caused by the imperfect experimental fidelity of state preparation at $W_d=0.7$, as shown in Table \ref{tab:shortcut_seq}.

\subsection{The transport process with the different lattice depth and acceleration}

Further, we study the influence of the lattice depth $V_0$ and acceleration $a_L$ on the transport process. To display the results, we define the atomic transport distance $S_a$ in half of one Bloch oscillation period with group velocity $v_g(t)$:
\begin{eqnarray}
	\label{Sa}
	S_a=\int_{t=0}^{T_B/2}v_g(t){\rm d}t,
\end{eqnarray}
where $T_B$ is the Bloch oscillation period. For $a_L=15 ~{\rm m/s^2}$, $T_B/2=287 ~{\rm \mu s}$.

\begin{figure}
	\includegraphics[width=0.45\textwidth]{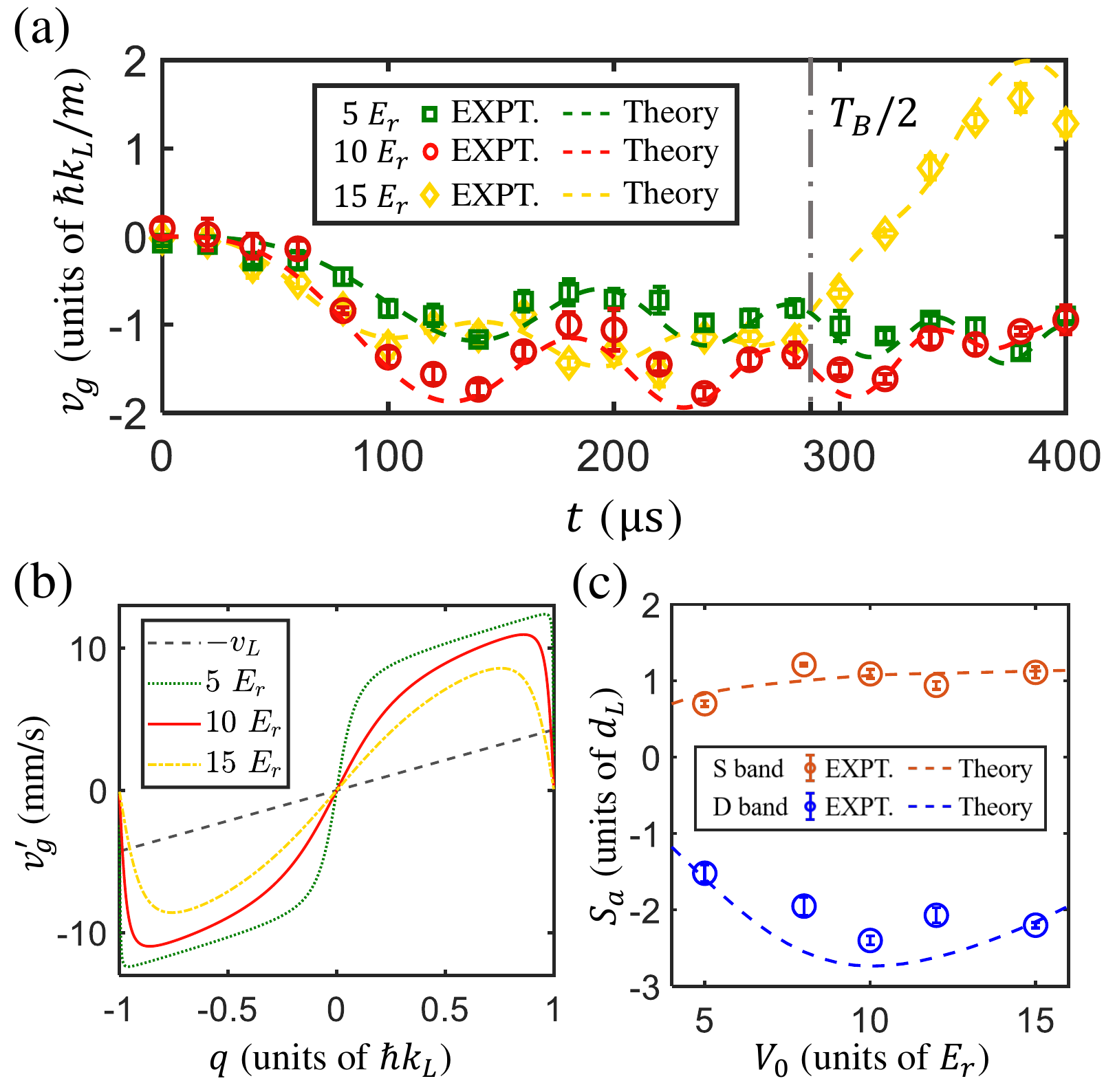}
	\caption{\textbf{Transport process of D-band atoms with the different lattice depth $V_0$.} 
		(a) The green squares, red circles and yellow diamonds represent the group velocity of D-band atoms in Lab frame with $V_0$=5,10,15 $E_r$, respectively. The dashed lines are the simulations. The dash-dotted line denotes half of one Bloch oscillation period $T_B/2$. The acceleration is $15 ~{\rm m/s^2}$.
		(b) The theoretical group velocity $v'_g$ of D band in moving frame with different lattice depth. The dotted green, solid red, and dash-dotted yellow lines denote the $v'_g$ with $V_0$=5,10,15 $E_r$, respectively. The dashed line denotes the velocity $-v_{L}=q/m$ of optical lattice. 
		In (c), the blue (lower) points and the orange (upper) points represent transport distance $S_a$ for atoms at D band and S band with $V_0$=5,8,10,12,15 $E_r$, respectively. The dashed lines are the simulations. $d_L=\lambda/2$ is the lattice constant. The errorbar denotes the standard error of five measurements.
	}\label{fig:Fig_result3}
\end{figure}

Fig.\ref{fig:Fig_result3}(a) shows the group velocities of D-band atoms with different $V_0$. The lattice depth $V_0$ is selected to keep the atoms in superfluid.
The transport process of D-band atoms is influenced by the change of D-band curvature and Landau-Zener tunneling.
On the one hand, Landau-Zener tunneling makes a part of atoms jump to higher bands and escape the optical lattice, which restrains Bloch oscillation. When $V_0$ is small, Bloch oscillation decays quickly, and the group velocity stops increasing before $T_B/2$, as the curve for $V_0=5~E_r$ shown in Fig.\ref{fig:Fig_result3}(a). 
As $V_0$ increases, Landau-Zener tunneling decreases, and the decay of Bloch oscillation reduces.
Because the curvatures on the opposite side of D band are reversed as shown in Fig. \ref{fig:Fig_principle}(a), when the decay of Bloch oscillation is small, the group velocity will reverse in the second half of the Bloch oscillations period, like the curve for $V_0=15 ~E_r$.
On the other hand, when the lattice depth increases, the change of D-band curvature makes its $v'_g$ decrease, as shown in Fig.\ref{fig:Fig_result3}(b).

Considering the two influences, with $V_0$ increasing, Landau-Zener tunneling decreases, and atoms at D band get more group velocity in half of the Bloch oscillations period, when the lattice depth is small. 
When the decay of Bloch oscillation in $T_B/2$ is small, the change of D-band curvature becomes important, and the group velocity obtained from the moving optical lattice decreases with $V_0$ increasing. Hence the absolute value of D-band atomic $S_a$ performs a trend to rise first and then fall with $V_0$ increasing, as shown in Fig. \ref{fig:Fig_result3}(c). The theoretical calculation is consistent with the non-monotonic behavior.

For atoms at S band, $v_g'$ is much lower than $-v_L=q/m$, so $v_g$ is dominated by $v_L$, as shown in Fig. \ref{fig:Fig_principle} (c). Because $v_L$ is independent of the lattice depth $V_0$, $S_a$ of S-band atoms is insensitive to $V_0$, as shown in Fig. \ref{fig:Fig_result3}(c).

\begin{figure}
	\includegraphics[width=0.36\textwidth]{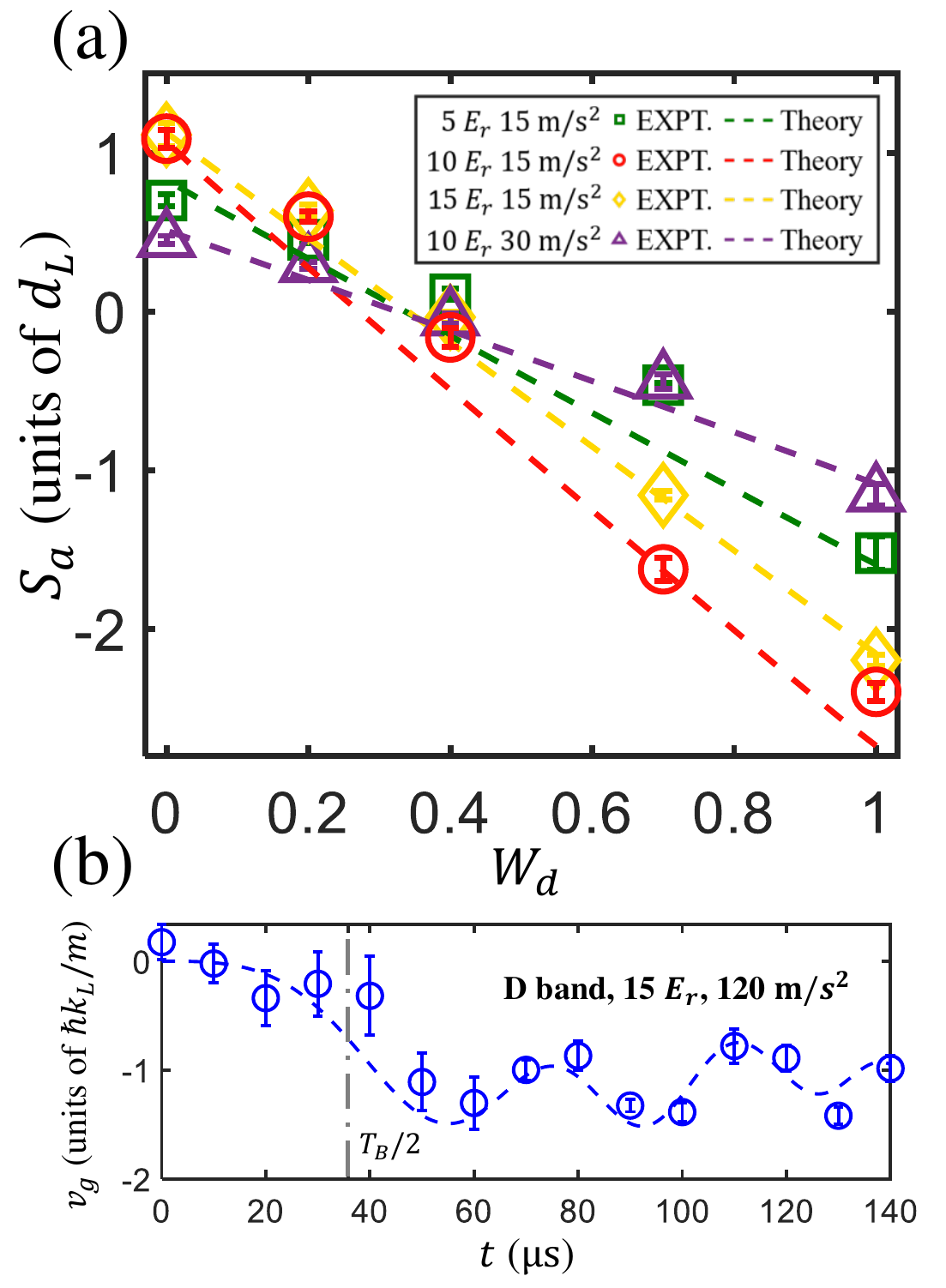}
	\caption{\textbf{Atomic transport distance $S_a$ with the different lattice depth $V_0$ and acceleration $a_L$.}
	(a)The green squares, red circles and yellow diamonds show transport distance $S_a$ with $V_0=5,10,15 ~E_r$, $a_L=15 ~{\rm m/s^2}$, respectively. The purple triangles show transport distance $S_a$ with $V_0=10 ~E_r$, $a_L=30 ~{\rm m/s^2}$.
	(b) shows the group velocity over time of D-band atoms in accelerated optical lattice when $a_L=120 ~{\rm m/s^2}$, when $V_0=10 ~E_r$. The dash-dotted line denotes half of one Bloch oscillation period $T_B/2$.
	The dashed lines are the simulations.
	The errorbar denotes the standard error of five measurements.
	}\label{fig:Fig_result4}
\end{figure}

Fig.\ref{fig:Fig_result4} (a) shows the transport distance $S_a$ with different $V_0$, $a_L$, and $W_d$, where the dashed lines denote the simulation results.
$S_a$ changes from positive to negative with $W_d$ increasing.
The changing rate of 10 $E_r$ is larger than that of 5 $E_r$ and 15 $E_r$, which is consistent with the trend of $S_a$ in Fig.\ref{fig:Fig_result3}(a).

Comparing to $V_0$, the change of $a_L$ only influences Landau-Zener tunneling, not the D-band curvature. Hence, with $a_L$ increasing, the atomic group velocity decreases, and $S_a$ reduces, as shown in Fig.\ref{fig:Fig_result4} (a).
From another point of view, when the acceleration is large, there is not enough time for atoms to respond to the moving optical lattice.
Fig.\ref{fig:Fig_result4} (b) shows the group velocity for $a=120 ~{\rm m/s^2}$, and in half of one Bloch oscillation period the atomic motion is small. Hence, $S_a$ reduces when $a_L$ increases.

\section{Discussion and Conclusions}
The discrepancy in group velocities between experimental results and theoretical simulations comes from several aspects.
Firstly, in the experiment, the error of lattice depth calibration is within $5\%$, and of acceleration calibration is about $2\%$. 
Secondly, due to the fluctuation of optical intensity and the imperfection of pulse waveform, the experimental fidelity of shortcut method is lower than the theoretical fidelity.
Thirdly, in the multi-orbital simulation, we ignore the atomic interaction and the momentum width of a condensate, which also influences the precision of simulations \cite{PhysRevLett.103.090403}. 
Moreover, the atoms on the excited bands will perform two-body collisions, and scatter to other bands \cite{PhysRevA.104.033326}. The collisions reduce the lifetime of superposition states, and influence the transport process.

Furthermore, the shortcut method can be used to load atoms into other bands or optical lattices with different spatial configurations, such as P and D band in one-dimensional and two-dimensional triangular lattice \cite{Zhou_2018,PhysRevLett.121.265301}, which helps to extend this protocol.
As for the excited bands, the transport process of P band in the moving lattice is also different from that driven by an external force in a stationary lattice.
$v_g'$ of P band in moving frame is lower than $-v_L$, like that of S band. So, in Lab frame, the atoms at S band and P band in a moving optical lattice will acquire the group velocities in the same direction.

In summary, using shortcut method, we observe the opposite group velocity of D-band and S-band atoms in a one-dimensional moving optical lattice. By the characteristic of atomic transport in the moving lattice, we perform the transport manipulation of superposition states with different superposition weights of D band and S band, where the quantum interference between atoms at different bands is observed.
Moreover, the influence of the lattice depth and acceleration on the transport process is studied.
The multi-orbital simulation method is used to calculate the transport process, and the calculations agree with our experimental results.
This study is helpful to understand transport phenomena at higher bands and detect topological properties \cite{science.abm6442}.

\section{Acknowledgement}
We thank Xiaopeng Li, Biao Wu, Yongqiang Li and Lingqi Kong for helpful discussion. 
This work is supported by the National Natural Science Foundation of China (Grants Nos. 11934002, 11920101004), the National Key Research and Development Program of China (Grant No. 2021YFA0718300, 2021YFA1400900), the Science and Technology Major Project of Shanxi (202101030201022), and the Space Application System of China Manned Space Program.

\appendix
\addcontentsline{toc}{section}{Appendices}\markboth{APPENDICES}{}

	

\bibliographystyle{apsrev}
\bibliography{my}

\end{document}